# Prediction of a gene regulatory network from gene expression Profiles with Linear Regression and Pearson Correlation Coefficient

Md Mehedi Hassan Onik
Department of CSE
Islamic University of Technology
Gazipur, Bangladesh.
onikrcc@gmail.com

Shakhawat Ahmmed Nobin
Department of CSE
Islamic University of Technology
Gazipur, Bangladesh.
shakhawat_ahmmed@yahoo.com

Adnan Ferdous Ashrafi
Department of CSE
Stamford University Bangladesh
Dhaka, Bangladesh.
nazib91@yahoo.com

Tareque Mohmud Chowdhury
Department of CSE
Islamic University of Technology
Gazipur, Bangladesh.
tareque@iut-dhaka.edu

***Abstract*— Reconstruction of gene regulatory networks is the process of identifying gene dependency from gene expression profile through some computation techniques. In our human body, though all cells pose similar genetic material but the activation state may vary. This variation in the activation of genes helps researchers to understand more about the function of the cells. Researchers get insight about diseases like mental illness, infectious disease, cancer disease and heart disease from microarray technology, etc. In this study, a cancer-specific gene regulatory network has been constructed using a simple and novel machine learning approach. In First Step, linear regression algorithm provided us the significant genes those expressed themselves differently. Next, regulatory relationships between the identified genes has been computed using Pearson correlation coefficient. Finally, the obtained results have been validated with the available databases and literatures. We can identify the hub genes and can be targeted for the cancer diagnosis.**

## I. Introduction

Sequencing the human genome is one of the important accomplishments in the history of System Biology. Human Genome contains valuable information about the traits, characteristics, expressions and also diseases of human beings. Cancer, the most dangerous one of them may be of various types has been the leading cause of death for recent years. According to WHO (World Health Organization) 8.2 million people died from cancer in 2012 and 20% of them could be cured if early detection would possible [1]. As no single gene decides how an organism grows therefore an understanding of gene regulatory network is the key that will open the door to those early detections of those diseases.

In this work we used Machine learning approach (Linear Regression based feature selection) to reduce the dimension of microarray dataset and Modified Pearson Correlation to reconstruct the GRN of [prostate] cancer. The cause of a disease is reflected in the change in expression level of genes.

Available sample number is very less in comparison to the total gene number [2]. It is very difficult to remove some genes since, we could have lost the important genes. Thus observed data contains a significant amount of noise [2]. Normally, both biological variations and experimental noise are the main cause for the difference in the measured transcript. To correctly interpret the gene expression in microarray data, it is crucial to understand the sources of the experimental noise [3].

Finally obtained network is much complex and large to represent and select the most significant genes [4].

It is difficult to select threshold for Pearson Correlation Technique. Since significant genes could be out due to bad selection of threshold. The threshold for which Maximum validate node is selected is chosen as threshold.

The main features of this work is to construct a gene regulatory network which will help biologist to identify the responsible cancer genes. We have proposed a combined approach which is robust comparing to others. The contributions of this work are:

- Handling the high-dimensionality problem by removing redundant genes without losing any significant information. These are done by measuring expression level of genes in different samples and get rid of those unwanted genes.
- We have reduced the complexity of analyzing the large microarray data.
- Our Proposed Method Used Pearson Correlation technique alone with linear regression.
- We constructed Network with different parameter with different amount of genes in different levels to show the actual network.
- For experimental result we have implemented our proposed method and figured out the performance using graphical and statistical approaches

Comparative analysis of our proposed method with different data set is given to show the strength of this combined technique.

## II. Related Works

### A. RELIEF-F

Relief-F is improved version of original Relief algorithm which has three important improvements and they are: less sensitivity to noise, better strategy for coping missing value and handling multiclass data [1].

### B. Wrapper and filter approach

Wrapper approach is wrapped within a learning algorithm. The filter approach is based on their information on dependency. Wrapper approach is used to predict the accuracy of a given feature subset. Some forward or backward selection algorithm is applied furthermore for better result. This process is repeated until the goal is achieved. But this method is expensive as it has the bigger run time. In bioinformatics most dataset are very large and it is largely used and then any classifier can be used for evaluate the classification accuracy of the test data which is not possible in wrapper approach.

### C. Spearman's rank correlation coefficient

Spearman's rank correlation coefficient is a nonparametric measure of statistical dependence between two variables. It describes the relationship between two variables using a monotonic function. Spearman's coefficient is appropriate for both continuous and discrete variables including ordinal variables. For a sample size n, the n raw scores $X_i$, $Y_i$ are converted to ranks $x_i$, $y_i$ and Þ is [5]:

$$Þ = 1 - \frac{0.6 \sum d_i^2}{n(n^2 - 1)} \tag{1}$$

where $d_i = x_i - y_i$ (difference between ranks)

### D. Bayesian network

BNs is linked with one more graphical model structure which is also termed as DAG (Directly Acyclic Graph). Mathematically rigorous and intuitively understandable are two noteworthy properties of BNs. Effective representation and computation of the joint probability distribution (JPD) over a set of random variables [6] are enabled by BNs. Two different ways are there to define Directed Acyclic Graph are: Set of Directed Edges and set of vertices [7]

### E. T test and fold change

T-statistic can be a reason of problem as the variance estimates due to genes shows a very low value for variance. These genes are associated to a large t-statistic and can be misjudged as differentially expressed [8]. If we apply it on small sample sizes which implies statistical power which is a great drawback. Thus, Using of T-test along with the importance of variance modeling may not be a reliable thing to use [9]. The fact the test was introduced more than 100 years ago should mean it is limited to some degree. The tests are based on limited theoretical assumptions and do not take into account all we know about these days. They are not specific over one sample, though it has been suggested over large samples their accuracy is approximately correct.

## III. Proposed Method

### A. Flowchart of our proposed method

Identifying Gene Regulatory Relationship between gene pairs with the help of gene expression profile, lots of different procedure have been used in the literature. In this work, for removing the redundant genes linear regression and for dependency among genes Pearson's correlation coefficient has been applied [11] [12].

### B. Removing Redundant Gene

Microarray dataset contains a large amount of redundant data, huge noise with very closely expressed value. Our main goal is to remove those closely expressed values, which ultimately gives no significance in gene selection through mean calculation of different genes expressions value. Our training dataset needed to be divide into two subtypes D1 and D2. Where calculation of the values for every gene in one subtypes is done. After that, compare those with another data types for all sample's average. This procedure applicable for all genes of the previous subtypes. We can easily remove those genes which gives similar values, as those will not give any discriminative information about gene's expression value.

### C. Linear Regression on Microarray Dataset

Defining an explanatory and a dependent variable is the key works in this regression analysis. We need to apply multiple target variable since our dataset contains multiple genes which is our target variable in this scenario. We need to compute Gn for each of the genes considering as a target variable and all other genes as dependent variable in the base subtype of the dataset to detect the regression model.

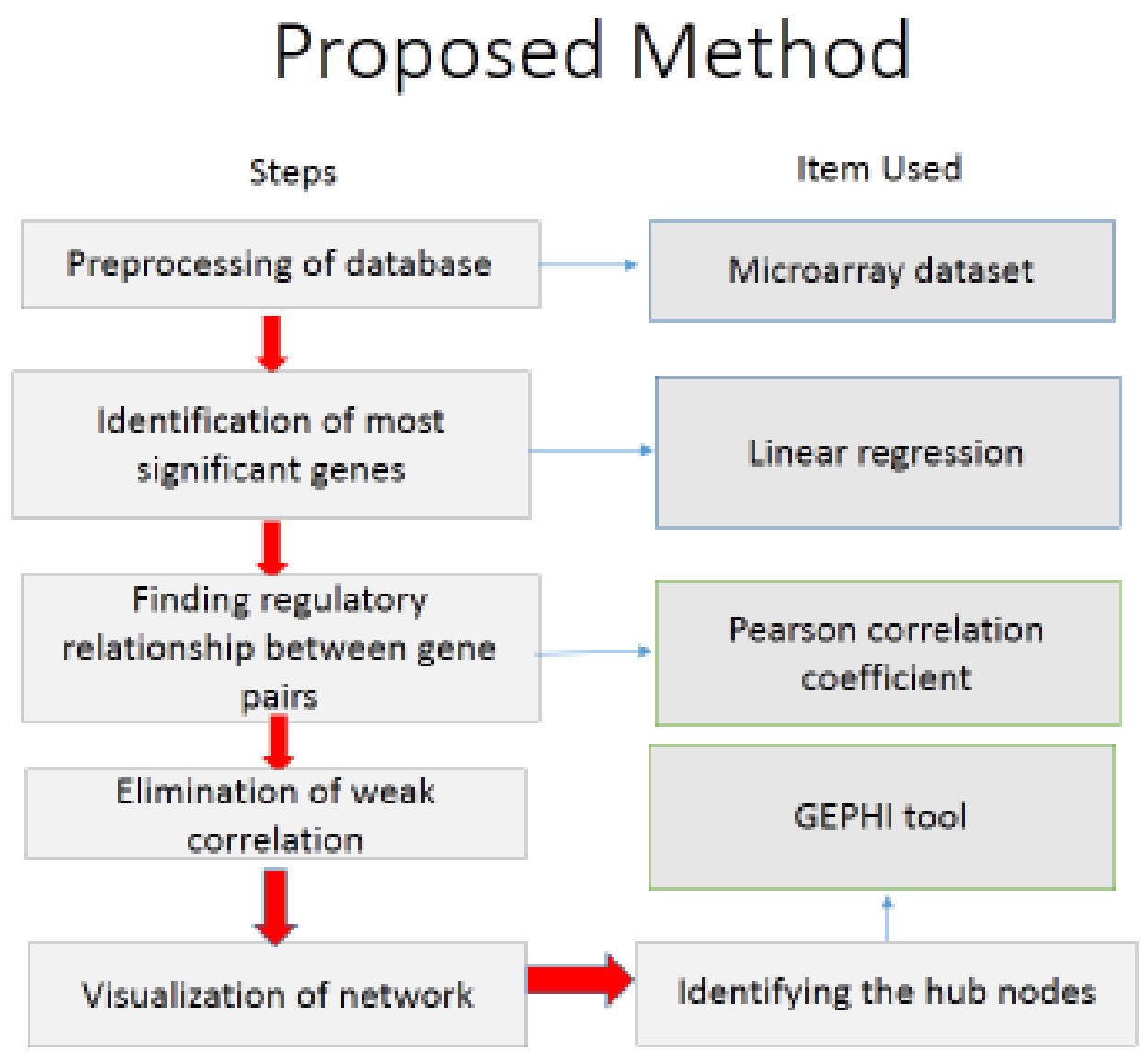

Fig. 1. Flowchart of our proposed method

$$G_1 = \beta_0 G_0 + \beta_2 G_2 + \beta_3 G_3 + \cdots + \beta_n G_n$$
$$G_2 = \beta_0 G_0 + \beta_1 G_1 + \beta_3 G_3 + \cdots + \beta_n G_n$$
$$\dots\dots\dots\dots\dots\dots\dots\dots\dots\dots\dots\dots\dots\dots\dots$$
$$G_n = \beta_0 G_0 + \beta_1 G_1 + \beta_2 G_2 + \cdots + \beta_{n-1} G_{n-1}$$

Equations stated above is the equation for linear regression model in which $G_n(i)$ denote an explanatory variable and other $g_j$ are the dependent variable without $g_i$. From gradient descent algorithm we can easily calculate parameter matrix (β) after considering all genes individually.

This β matrix represents the regression model for the subtype of dataset that has been considered for comparison with the other subtype, where each row of β ($\beta_i$) is the set of parameters for a particular $G_n(i)$. Using the transpose of this matrix we, statistically can predict the gene expression values by applying $\beta_{(i)}$ on the other subtype of the training dataset and calculate the $G_n'(i)$. This is done by equation where $g_i'$ is the feature vector of the second subtype of dataset and is $\beta_i^T$ transpose of parameter vector generated from the first subtype of dataset.

$$G_n' = \beta_i^T g_i' \qquad (2)$$

Our model was designed in such way where we divide our actual dataset into two parts: training dataset and test dataset. From the training dataset the two different subtypes of data have been separated.

*1) Algorithm 1:*

INPUT: $D_1$ and $D_2$ are two subtypes of training dataset. N is the number of features and the number of samples is M.

OUTPUT: A significant set of features.

1. Mean values for $D_1$ and $D_2$ on every features needed to calculate
2. Find the difference of mean values between $D_1$ and $D_2$ and sort features on them
3. Remove the features with smaller difference (no discriminant expression values)
4. Apply linear regression analysis where each features as predictive variable
5. Calculate Parameter β for each features
6. Create a parameter matrix (β) of size (n×n) for $D_1$
7. Calculate $G_n'(i)$ value from Equation (2)
8. Calculate the divergence of expression values from standard and sort the features based on their divergence value

The divergence of genes expression value means those features got some change from ideal value. We have ranked the genes; those are selected based on their divergence from the standard model of regression line. The most deviated gene from the regression line got the highest rank in the feature subset list. The more the expression value is diverged from the regression line, the better possibility the feature is a discriminative feature.

### *D. Identifying Regulatory Relationship*

To create a regulatory relationship, we must bring a relation among gene pairs. We proposed the Pearson correlation technique to detect how much related the most significant genes are that we get from our previous steps.

Correlation means sets of data measuring to detect how well they are related. This is a statistical technique where Pearson correlation is chosen for significant result outcome. It shows linear relationship between two data. We calculate the value 'r' to detect how strongly they are bonded. We follow this equation to calculate r:

$$r = \frac{n(\sum xy) - (\sum x)(\sum y)}{\sqrt{[n\sum x^2 - (\sum x)^2][n\sum y^2 - (\sum y)^2]}} \qquad (3)$$

Where n is the total number of genes, and x and y are two genes between which genes we will calculate the r value.

*1) Algorithm 2:*

INPUT: A set of significant genes
OUTPUT: Weighted Relation between all those genes.

1. Take one gene and find r value with all other genes according to equation 3
2. Do 1 for all genes to calculate r for all genes
3. Calculate the absolute r value which is greater than .85

The result of the Pearson Correlation is between -1 to 1 though it's rare to get -1, 0 or 1. Three main types of correlation is present those are change like this

**High Correlation**: 0.5 to 1.0 or -0.5 to 1.0
**Medium Correlation**: 0.3 to 0.5 or -0.3 to 0.5
**Weak Correlation**: 0.1 to 0.3 or -0.1 to 0.3

We have selected highly correlated genes only in our work to get largely dependent gene pair.

### *E. Construction of Gene Regulatory Network*

Now gene interaction network has been constructed, where nodes correspond to gene names and pairwise r value is allocated to the edge between genes. We use a software name gephi to construct network diagram.

### *F. Identifying Genes Responsible for Cancer*

Genes are treated as Node and relation between them as edge. Therefor the nodes with most degree is related with most genes. That means those are treated as hub node which can be taken into action for early detection or medicated action to take.

## IV. Result Analysis

### *A. Data Set and Experimental Setup*

In our experiment, we have calculated the relationship among genes in 'Matlab 2013' and construction of Gene Regulatory Relationship (GRN) with the 'Gephi' graph design tool. All the simulation are performed on a personal computer of 2.13GHz processors with 2 GB main memory. Here we have used 3 data set.

TABLE I. DATASETS USED

| Data Type | Class | Samples | Genes | Purpose of Use |
|---|---|---|---|---|
| All Data | 2 Class(one cancer positive & one cancer negative) | 128 | 12625 | Stability test |
| Colon Tumor Data 1 | 2 Class(one cancer positive & one cancer negative) | 62 | 2000 | Stability test |
| Colon Tumor Data 2 | 2 Class(one cancer positive & one cancer negative) | 20 | 15552 | Accuracy and Validation |

Among them two (one all and one colon) are used for checking the strength of the proposed procedure and last dataset (colon) used for validation of our result with accuracy.

## B. *Performance Analysis*

In our study we checked the strength of the proposed procedure and then use to find genes responsible. After getting a set of responsible genes we matched those with validated gene set to get result and accuracy.

### 1) *Stability checking of Proposed Procedure*

In our dataset there were total 62 samples. Among them 19 sample was cancer positive and 41 was cancer negative. We divided those into two subgroups and apply our procedure parallels into two sub group to find the result obtained. We have seen that our method was detected almost same genes from subgroup 1 and subgroup 2, which proved stability of our method.

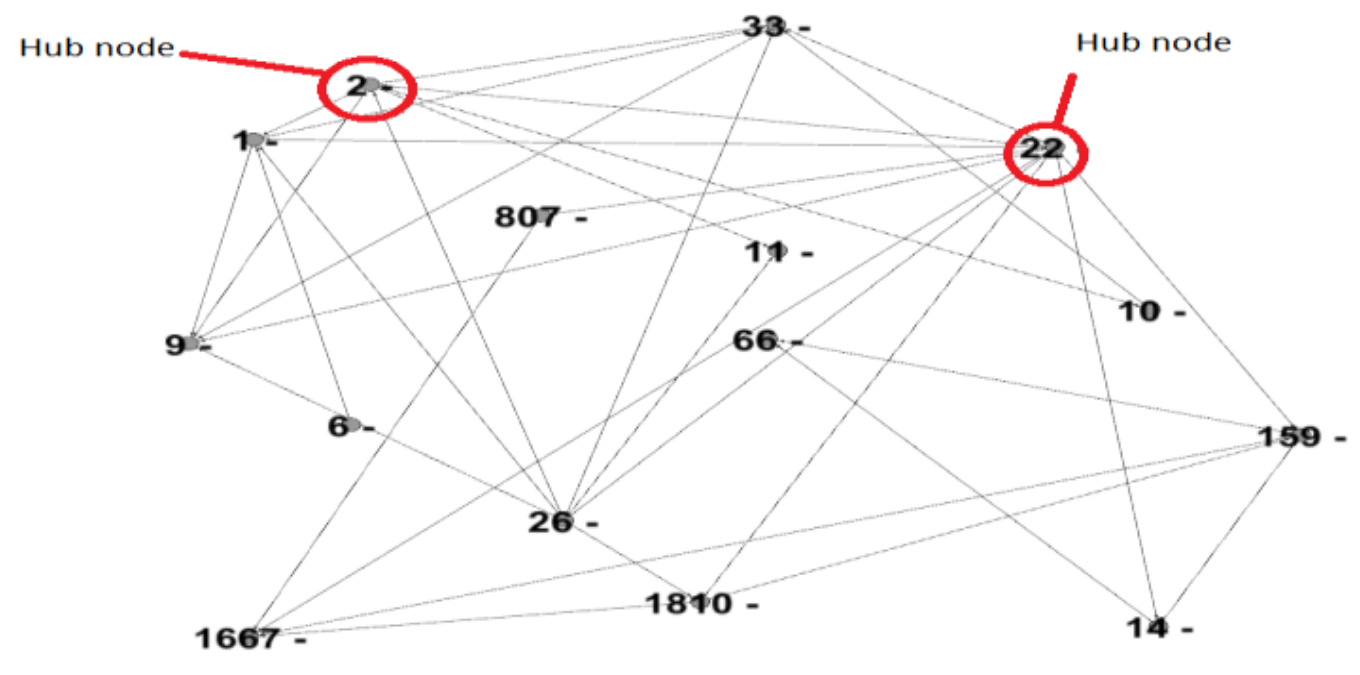

Fig. 2. Identifying Genes Responsible for Cancer

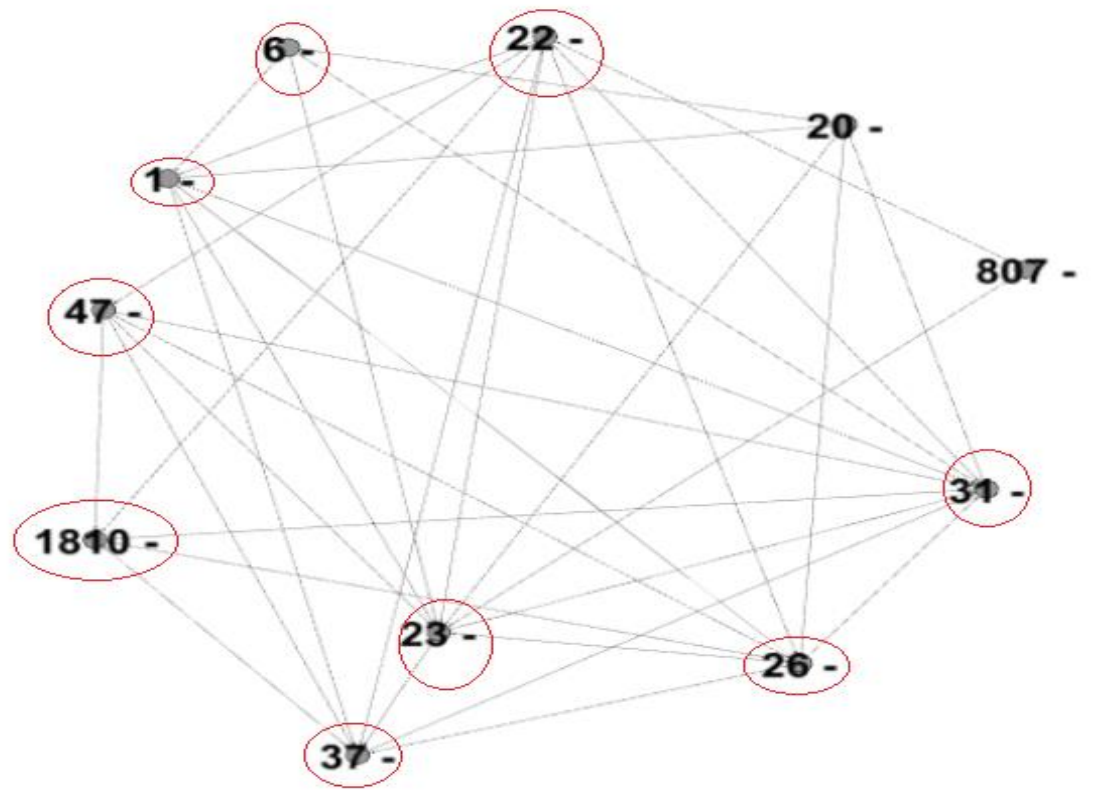

Fig. 3. Comparison analysis Subgroup1 (30 genes)

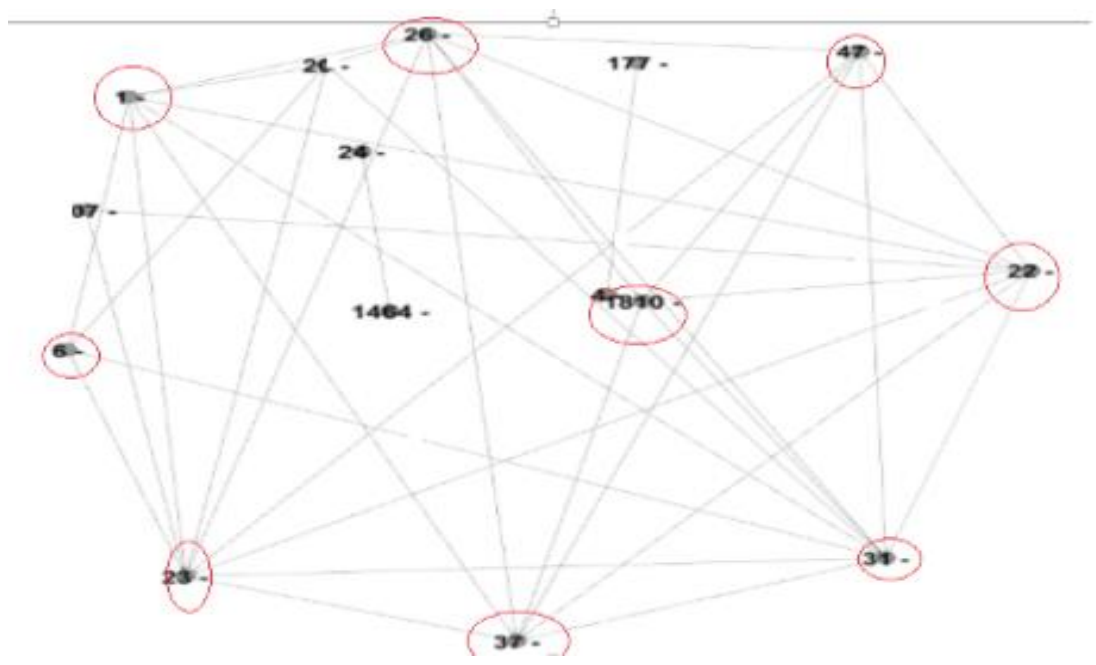

Fig. 4. Comparison analysis Subgroup2 (30 genes)

In our study, we divide the data set in two part. Individually we apply our method on both. Then we compare each and check if there is any common set of gene in two parts or not. We find a common set of genes having in both result. That satisfy the stability of our method. In table we will show the comparison.

### 2) *Validation and Accuracy:*

From literature review and available dataset have got some genes like APC, MUTYH, TP, EPCAM, BMPR are responsible for cancer in a sample [13]. And 5 of them are present in our dataset we took 100 genes for our procedure and 3 of them were also present in our resultant constructed genes shown in table 4.2.

TABLE II. SIMILARITY BETWEEN TWO SUBGROUP

| Gene ID | Group-1 | Group-2 |
|---|---|---|
| 1 | ☑ | ☑ |
| 6 | ☑ | ☑ |
| 20 | | ☑ |
| 22 | ☑ | |
| 24 | ☑ | ☑ |
| 37 | ☑ | ☑ |
| 97 | ☑ | |
| 1464 | ☑ | |
| 1810 | ☑ | ☑ |

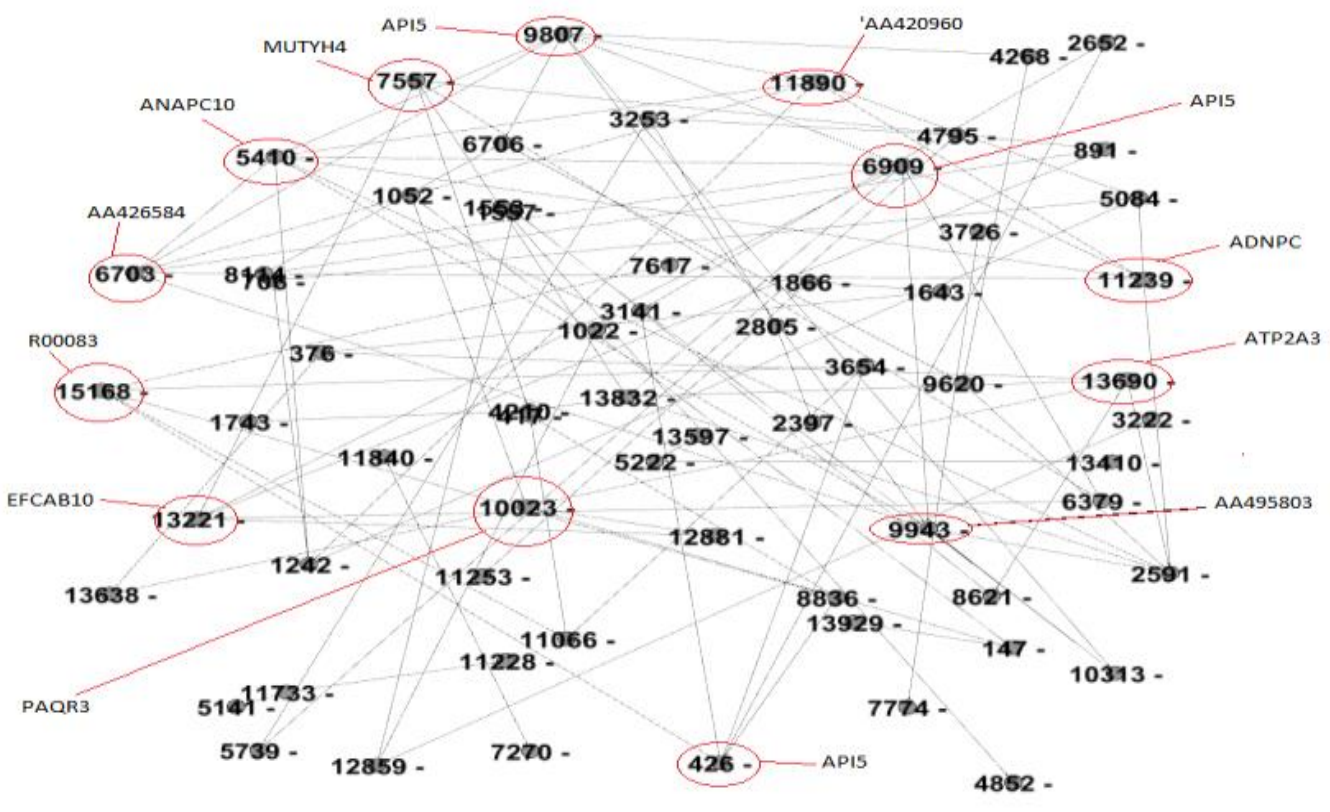

Fig. 5. Final GRN with 100 genes from dataset-1

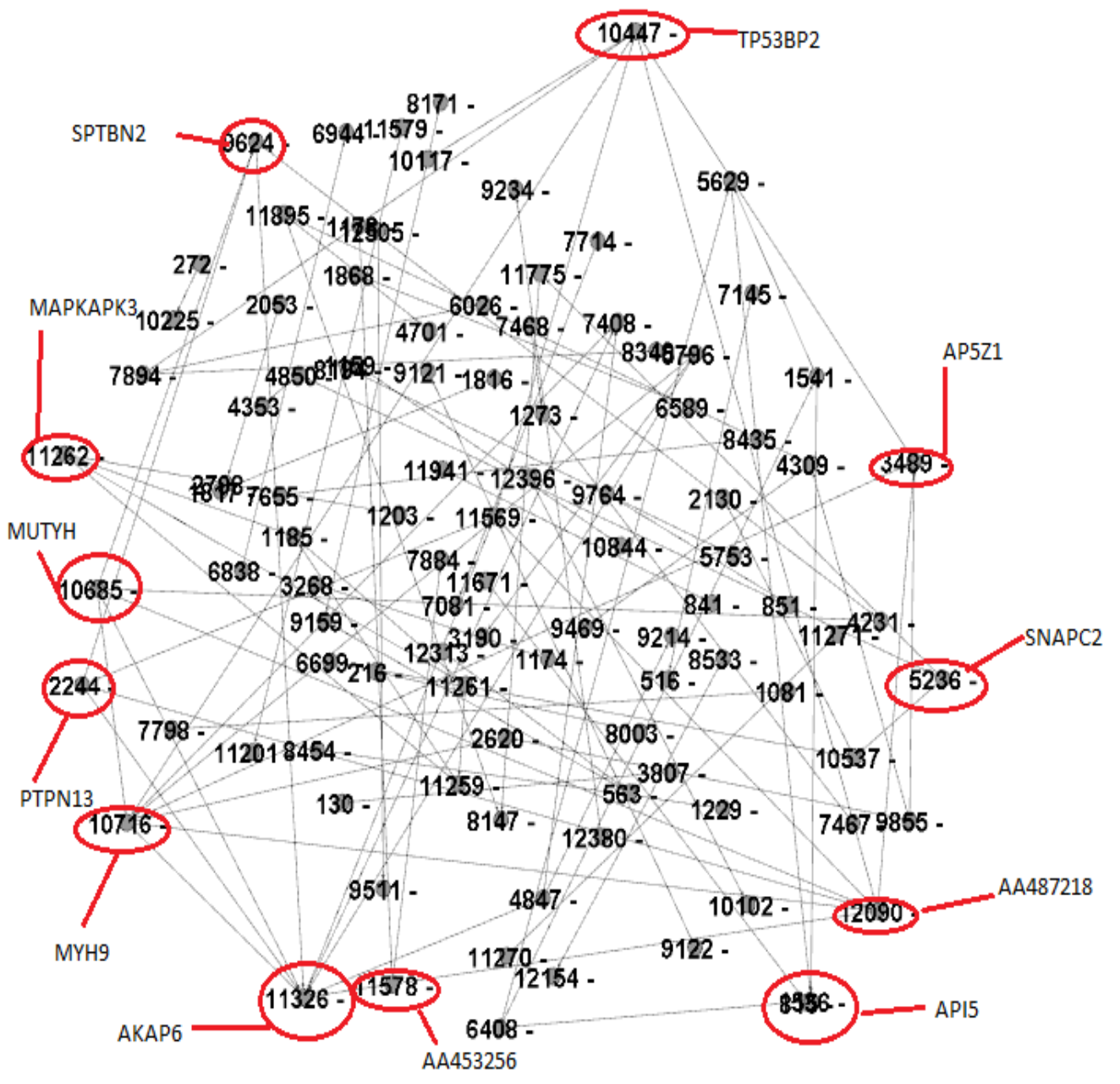

Fig. 6. Final GRN with 100 genes from dataset-2

From the above table we can see the genes which are mostly responsible and now we will mark those genes which are also available in our study. We applied the method in two datasets. Common genes are marked with red circle alone with their number of degree for which they are treating as responsible genes.

Accuracy = (Detected Genes by our Procedure) / (Actual Genes Detected from literature)

$$Accuracy = \frac{3}{5} = 0.60$$

By using this equation, we calculate our estimated accuracy where. From literature we APC, MUTYH, TP, EPCAM, BMPR got genes responsible for cancer in our dataset and among them APC, MUTHY, TP Were present [14]. Finally, we got 3 Genes which match with the genes from those literature and publications. Thus lastly we got accuracy of 60% on a particular dataset.

| Serial No. | Gene ID | Genes name | Top genes with degree |
|---|---|---|---|
| 1. | 9807 | API5 | 8 |
| 2. | 6909 | SIN3B | 7 |
| 3. | 13690 | ATP2A3 | 7 |
| 4. | 10023 | PAQR3 | 7 |
| 5. | 9943 | AA495803 | 6 |
| 6. | 6703 | AA426584 | 6 |
| 7. | 426 | API53 | 6 |
| 8. | 13221 | EFCAB10 | 5 |
| 9. | 15168 | R0083 | 5 |
| 10. | 7557 | MUTYH4 | 5 |
| 11. | 11890 | AA420960 | 5 |
| 12. | 13690 | ATP2A3 | 5 |
| 13. | 11239 | ADNPC | 5 |

Fig. 7. Highly connected genes involve in network construction for dataset 1

| Serial No. | Gene ID | Genes name | Top genes with degree |
|---|---|---|---|
| 1. | 11326 | AkAP6 | 7 |
| 2. | 10716 | MYH9 | 7 |
| 3. | 12090 | AA487218 | 6 |
| 4. | 10447 | TP53BP2 | 6 |
| 5. | 106685 | MUTYH | 6 |
| 6. | 11578 | AA53256 | 5 |
| 7. | 9624 | SPTBN2 | 4 |
| 8. | 13221 | PYPN13 | 4 |
| 9. | 3489 | AP5Z1 | 4 |
| 10. | 8356 | MUTYH4 | 4 |
| 11. | 11890 | AA487218 | 4 |
| 12. | 11262 | MAPKAPK3 | 4 |
| 13. | 2244 | PTPN | 4 |

Fig. 8. Highly connected genes involve in network construction dataset 2

### 3) *Comparison Analysis*

A number of approaches has been followed to identify the genes responsible for cancer. Here this study we will compare only those approaches which is similar with our technique. In the table below we have shown the final genes identified by other method and our method. Here we compared the result with the names found in the different literature and website (biological database). We have shown result of two datasets.

TABLE III. COMPARISON AMONG LITERATURE REVIEWED GENES AND OBTAINED GENE

| **Name of Genes Found in literature and Biological Database** | **T-test + Fold change with Pearson Correlation** | **T-test + Fold change with Mutual Information** | **Linear Regression with Pearson Correlation (Dataset 1)** | **Linear Regression with Pearson Correlation (Dataset 2)** |
|---|---|---|---|---|
| APC(Colon) | | | √ | √ |
| TP53(Multiple) | | √ | √ | √ |
| MLH1(Multiple) | | | | |
| MUTYH(colon) | | | √ | √ |
| POLD1(Colon) | √ | √ | √ | |
| ACAT2(Multiple) | √ | √ | | √ |

## V. Conclusion

### A. Summary of Contribution

The complex molecular interaction is for perturbation in the GRN. So, detecting the cancerous genes is a key step for cancer diagnosis. A regulatory network give idea among genes interactions and dependency. In our procedure we took a machine learning approach to find the most significant genes, Pearson correlation between gene-pairs to reconstruction of gene regulatory network. Where we took a sample of 15552 genes and from there after linear regression analysis we got 100 most significant genes after that we apply Pearson correlation model on those 100 Genes to get Pearson factor ‘r’ value. From those genes we only consider genes which got r >.85 and get 56 correlated genes. After reconstruction of those gens we found 12 as a hub nodes and from literature we have found 3 are similar with our procedural result output.

### B. Limitation and future work

Due to difficulty in dataset availability, the construction of gene regulatory networks and their validation in a realistic manner is really a difficult task. Our proposed approach can help to identify common molecular interaction in the cancer study not only in colon cancer but other cancer like lung, breast, etc. In future we will try to implement with other dataset for construction of those types of cancer.